\documentclass[useAMS,usenatbib]{mn2e}
\usepackage{amsmath}
\usepackage[toc,page]{appendix}
\usepackage{pslatex}
\usepackage{amsmath}

\usepackage{graphicx}
\usepackage{float}

\title[Thermal surface waves in a protoplanetary disk]{Simulation of Thermal Surface Waves in a Protoplanetary Disk
in a Two-Dimensional Approximation}
\author[Pavlyuchenkov et. al.]{Ya. N. Pavlyuchenkov, L. A. Maksimova, V. V. Akimkin \\
        Institute of Astronomy RAS, Moscow, Russia}
\date{}
\begin{document}
\date{Received 10.06.2022; revised 11.08.2022; accepted 30.08.2022 \\
pavyar@inasan.ru}


\maketitle

\begin{abstract}
\noindent
Theoretical models predict that the obscuration of stellar radiation by
irregularities on the surface of a protoplanetary disk can cause
self-generating waves traveling towards the star. However, this process
is traditionally simulated using the 1+1D approach, the key
approximations of which~--- vertical hydrostatic equilibrium of the disk
and vertical diffusion of IR radiation~--- can distort the picture. This
article presents a two-dimensional radiative hydrodynamic model of the
evolution of an axially symmetric gas and dust disk. Within this model,
but using simplified assumptions from 1+1D models, we have reproduced the
spontaneous  generation and propagation of thermal surface waves. The key
conclusion of our work is that taking into account two-dimensional
hydrodynamics and diffusion of IR radiation suppresses the spontaneous
generation and  development of thermal waves observed in the 1+1D
approximation. The search for the possibility of the existence of surface
thermal waves should be continued by studying the problem for various
parameters of protoplanetary disks. 
 \vspace{0.2cm}
\newline {\bf DOI:} 10.1134/S1063772922100110
\vspace{0.3cm}
\end{abstract}

\section{Introduction}
In protoplanetary disks, conditions for the emergence  of a wide variety
of dynamic instabilities are realized, the development of which can
affect both the observed manifestations and the overall evolution of
disks 
\citep{Velikhov1959,1991ApJ...376..214B,2005ApJ...620..459Y,2013MNRAS.435.2610N,2014ApJ...788...21K,2016MNRAS.462.4549L,2022MNRAS.512.2636Z}.
One of such instabilities is that associated with the obscuration of
stellar radiation by the surface inhomogeneities of the disk
\citep{2008ApJ...672.1183W}. This instability is due to the positive
feedback between the angle at which the star’s radiation enters the
disk’s atmosphere and its heating. Many theoretical models show that a
small local distortion of the disk surface can provoke the generation of
waves traveling towards the central  star,
see~\cite{2021ApJ...914L..38U,2021ApJ...923..123W,2022arXiv220109241O}.
Such thermal waves on the disk surface were predicted both in the inner
regions of passive disks, where the characteristic time of establishment
of thermal  equilibrium is shorter than the dynamic
time~\citep{2008ApJ...672.1183W},  and on the disk periphery in the
reverse situation \citep{2000A&A...361L..17D}. In our previous
work~\citep{2022ARep...66..321P}, we also reproduced the  process of
formation of thermal waves and found that they can affect only the upper
layers of disks without significant  temperature fluctuations in the
equatorial plane.

However, the 1+1D approach used by~\cite{2022ARep...66..321P},  as in
most other works on this problem, is based on several key approximations
that can significantly distort the real picture. Such approximations are
1) the absence of diffusion of thermal radiation in the radial direction
and 2) hydrostatic equilibrium in the vertical direction and the absence
of gas-dynamic effects in the radial direction. The gradual abandonment
of these assumptions is important for substantiating the reality of
thermal waves in real disks.

In particular, in recent paper by~\cite{2022arXiv220109241O}, a two-layer
disk model was presented, in which two-dimensional ($R,z$) effects of
heating the equatorial layers by IR radiation  from surface layers were
considered and the development of instability for various parameters of
the numerical model was demonstrated. However, considering the diffusion
of IR radiation in the radial direction in even more detail can lead to a
wide  redistribution of thermal energy in the vicinity of the
perturbation,  which can smooth out the resulting waves.

\begin{figure*}
\includegraphics[width = 0.4\linewidth]{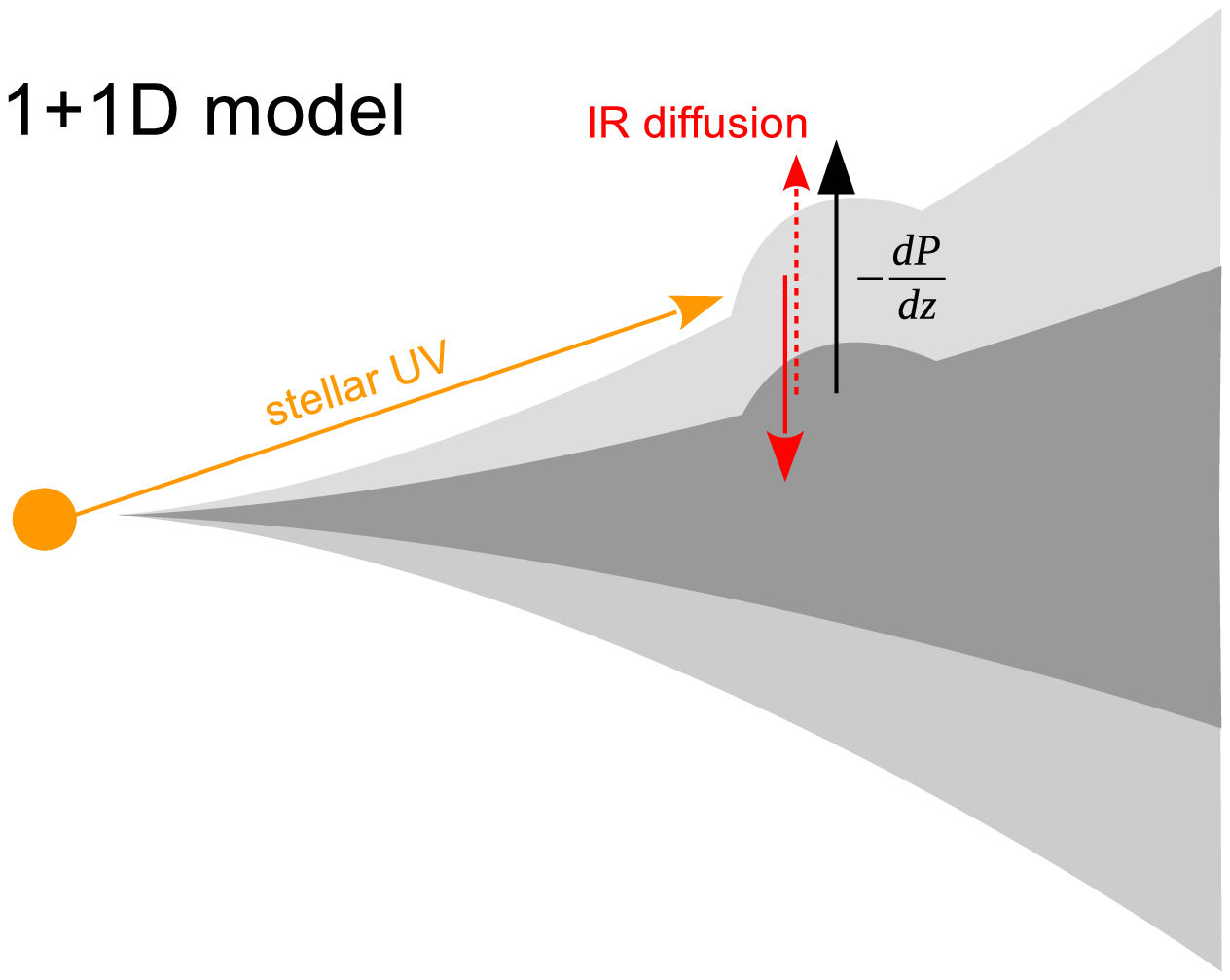}\hspace{0.5cm}
\includegraphics[width = 0.4\linewidth]{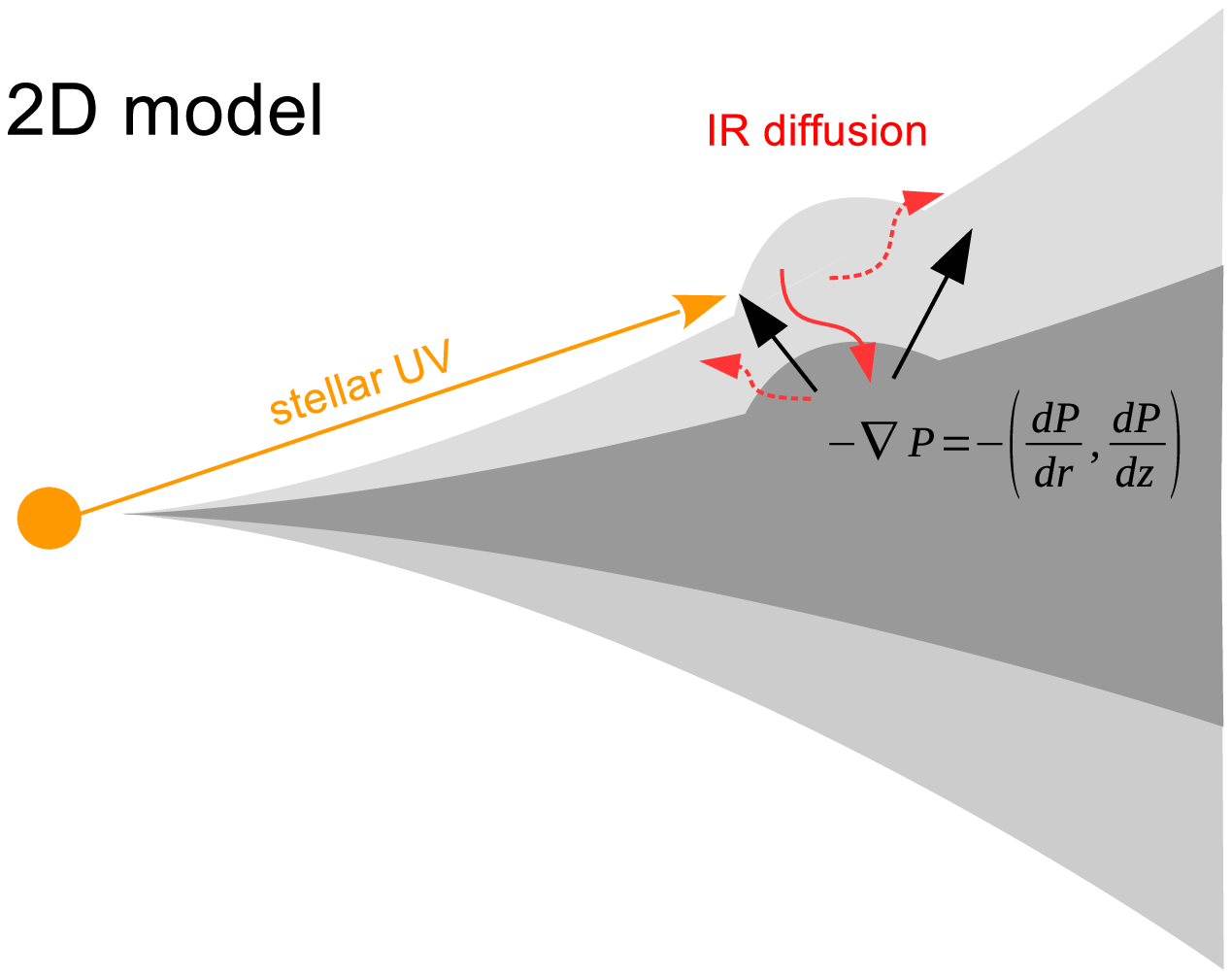}
\caption{Illustration of the processes during the formation of a hump on
the disk surface (left) in the 1+1D model and (right) in the 2D model:
(black arrows) the forces associated with the pressure gradient, (red
solid arrows) IR radiation that heats the inner layers of the disk, and
(red dotted arrows) outgoing IR radiation.}
\label{intro}
\end{figure*}

The initial perturbation can also be reduced or even completely
suppressed by dynamic effects. Indeed, increased pressure in the heated
layer can be used not only to lift the layer (as in the 1+1D model), but
also to push apart neighboring radial layers. A schematic of these
processes is shown in Fig.~\ref{intro}.

The aim of this work is to consider the processes of gas dynamics and
diffusion of IR radiation in the excitation  of surface thermal waves.
This study is conducted  using a fully two-dimensional radiative
hydrodynamic  model. Within this model, we also implemented the 
simplifying assumptions of the 1+1D approaches in order to determine
their validity.

\section{Basic Axisymmetric Model of a Protoplanetary Disk}
\label{sec:base_model}

\begin{figure*}
\includegraphics[width = 0.49\linewidth]{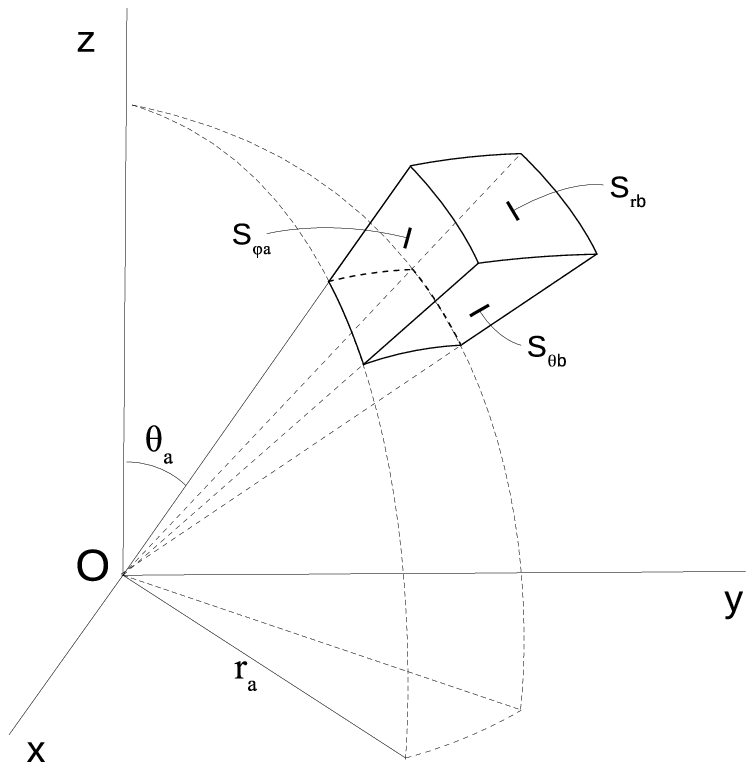}
\includegraphics[width = 0.49\linewidth]{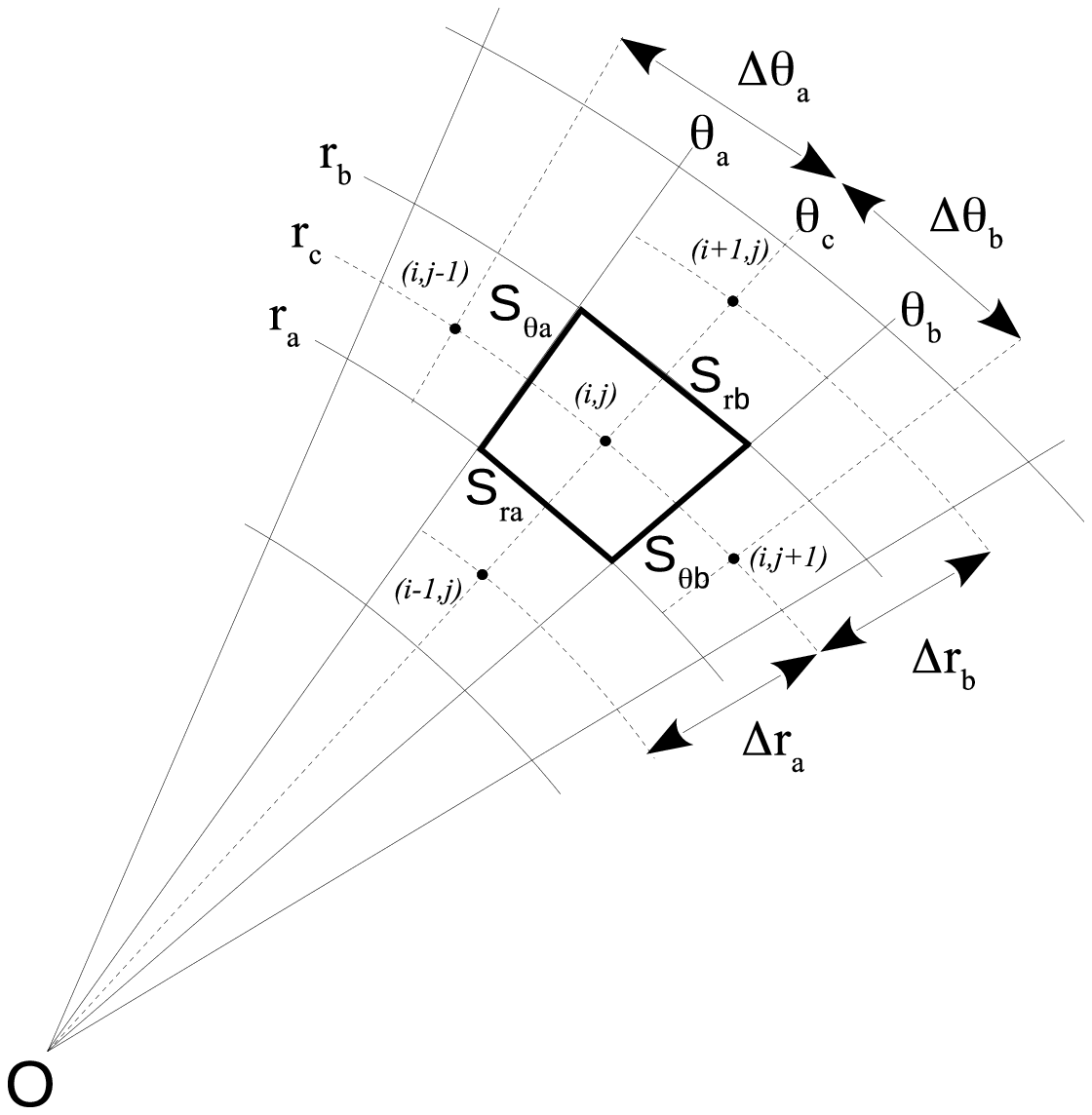}
\caption{Structure of the computational grid with the designations  of
quantities: (left) an individual cell in spherical coordinates and
(right) polar cut of the grid.}
\label{scheme1}
\end{figure*}

To simulate the evolution of a gas and dust disk, we use a combination of
finite difference methods for hydrodynamics and radiative transfer
adapted to a spherical coordinate system. The entire computational domain
is divided into cells, within which the values of physical quantities are
assumed to be constant. The grid structure is shown in
Fig.~\ref{scheme1}. We consider the spherical coordinate system more
convenient than the others, because the calculation of the heating of the
medium by the UV radiation of the star (ray tracing in the radial
direction) in the spherical coordinates is implemented most easily. Axial
symmetry in spherical coordinates is implemented by introducing a single
cell in the $\varphi$ coordinate. In our calculations, we use an
inhomogeneous discrete grid refined towards the center  in the radial
($r$) direction and towards the equator when splitting in the angle
$\theta$ with a resolution of 360 radial $\times$ 64 angular cells. The
cell length in the angle $\varphi$ is taken equal to one degree, which is
comparable  to the resolution in $\theta$ near the equator. Such a grid
reflects well the disk structure and makes it possible  to trace the
resulting gradients of physical quantities.  The evolution is calculated
using the splitting with respect to physical processes; i.e., the
hydrodynamic step is followed by the radiative transfer calculation step.

\subsection{Hydrodynamic Method}
To describe the dynamic evolution of a gas and dust disk, we use the
standard gas dynamics equations for an inviscid gas in divergent form:
\begin{eqnarray}
&&\frac{\partial \rho}{\partial t} + {\nabla} \cdot (\rho {\bf U}) = 0 \\
&&\frac{\partial }{\partial t}(\rho {\bf U}) + {\nabla} \cdot (\rho {\bf U U}+P) = \rho {\bf f} \\
&&\frac{\partial E}{\partial t}+ {\nabla} \cdot ({\bf U}(E+P)) = \rho {\bf f} \cdot {\bf U},
\end{eqnarray}
where $\rho$ is the bulk density, ${\bf U}$ is the velocity, $P$ is 
pressure, ${\bf f}$ is the gravitational force per unit mass, $E =
\dfrac{P}{\gamma-1} + \dfrac{\rho U^2}{2}$ is the total energy of the gas
per unit volume, and $\gamma$ is the adiabatic exponent.

To solve this system, we use the classical Godunov method, a detailed
description of which can be found in~[\cite{Kulikovskii:book}, Section
3]. In this method, gas-dynamic fluxes through cell boundaries are found
as a result of solving the problem of the decay of an arbitrary
gas-dynamic discontinuity. In the implementation we use, the 
discontinuity decay problem is solved exactly using the bisection method
for the resulting nonlinear equation. The found fluxes between cells are
used to calculate the physical quantities in the cells on a new time
layer. The finite-difference scheme was implemented within the formalism
described in~\cite{Abakumov:2014}, where Godunov-type difference schemes
in curvilinear coordinates, their application in spherical coordinates,
and test examples are considered.

Ultimately, the difference scheme we implemented reduces to the following
calculations. Let ${\bf q} = (\rho, \rho u, \rho v, \rho w, E)^{T}$  be a
vector of conservative variables,  where $\rho$ is the bulk density and
$u$, $v$, and $w$ are the velocity components in the spherical
coordinates. To find the value of ${\bf q}^{n+1}$ on a new time layer,
two steps are performed: the advection step and the step of  considering
gravitational sources. The advection step is reduced to finding
intermediate values ${\bf q}^{*}$ as follows:
\begin{equation}
{\bf q}^{*} = {\bf q}^{n} +
\frac{\Delta t}{\Delta V} \left(\Delta{\bf F} + \Delta{\bf G} + \Delta{\bf H}\right),
\label{advection}
\end{equation}
where $\Delta t$ is the time step; $\Delta V$ is the volume  of the
current cell; and $\Delta{\bf F}$, $\Delta{\bf G}$,  and $\Delta{\bf H}$
are the fluxes through the corresponding cell faces, the components of
which are given in Appendix~\ref{AppA}. The advection step is  followed
by the step of taking into account gravity sources, at which the
correction for the radial velocity is calculated:
\begin{equation}
u^{n+1} = u^{*} - \dfrac{GM}{r^2} \Delta t,
\end{equation}
where $G$ is the gravitational constant, $M$ is the mass of the central
star, and the quantities dependent on it are recalculated, forming the
${\bf q}^{n+1}$ values on the new time layer:
\begin{equation}
{\bf q}^{n+1} = (\rho^{*}, \rho^{*} u^{n+1}, \rho^{*} v^{*}, \rho^{*} w^{*}, E(u^{n+1},v^{*},w^{*}))^{T}.
\end{equation}

The hydrodynamic method implemented was thoroughly tested. In particular,
the numerical  solution of the discontinuity decay problem agrees well
with the analytical solution for all types of  discontinuities. In
addition, we made sure that, within the  one-dimensional axisymmetric
geometry, the finite-difference  approach used gives close results in
comparison with the classical approach, in which the source terms
associated with the curvilinearity of the coordinate system are clearly
distinguished.

\subsection{Method for Calculating the Radiative Transfer}

To calculate the thermal structure of a gas and dust disk, we use a
generalization of the nonstationary  thermal model
from~\cite{2017A&A...606A...5V} to the two-dimensional case. The model
takes into account the heating of the medium by the direct radiation of
the star and diffusion of  thermal radiation. The corresponding system of
equations has the form
\begin{eqnarray}
\rho c_{\rm V} \frac{\partial T}{\partial t}&=& c \rho \kappa_{\rm P} \left(E_{\rm r} - a T^4\right) + s_{\star}
\label{therm_sys1}\\
\frac{\partial E_{\rm r}}{\partial t}&=& - c \rho \kappa_{\rm P} \left(E_{\rm r} - a T^4\right) + \hat{\Lambda} E_{\rm r},
\label{therm_sys2}
\end{eqnarray}
where $\rho$ is the density of the gas and dust medium, $c_{\rm V}$ is
the specific heat of the medium [\,erg g$^{-1}$ K$^{-1}$], $c$ is the
speed of light, $\kappa_{\rm P}$ [\,cm$^2$ g$^{-1}$] is the
Planck-averaged IR radiation true absorption coefficient (without the
contribution of scattering, per unit mass of the gas and dust medium),
$s_{\star}$ [\,erg cm$^{-3}$ s$^{-1}$]  is the rate of heating by stellar
radiation, $T$ is the medium temperature, and $E_{\rm r}$ is the energy
density of IR radiation. Equation \eqref{therm_sys1} describes the change
in the volumetric thermal energy of the medium as a result of absorption
and re-emission  of thermal IR radiation (the terms $c\rho\kappa_{\rm
P}E_{\rm r}$ and $c\rho\kappa_{\rm P}a T^4$, respectively), as well as a
result of absorption of star’s direct UV radiation ($s_{\star}$).
Equation \eqref{therm_sys2} is a moment equation of radiative transfer in
the Eddington  approximation and describes the change in the energy
density of IR radiation as a result of absorption and re-emission of
thermal IR radiation, as well as a result of spatial diffusion of IR
radiation, represented by the operator $\hat{\Lambda}E_{\rm r}$:
\begin{equation}
\hat{\Lambda}E_{\rm r} = -{\rm div}\, {\bf F}_{\rm r} = {\rm div} \left( \frac{1}{\sigma} {\rm grad}\, E_{\rm r} \right),
\label{operE}
\end{equation}
where ${\bf F}_{\rm r}$ is the IR radiation flux, $\sigma =
3\rho\kappa_{\rm R}/c$, and $\kappa_{\rm R}$ [\,cm$^2$ g$^{-1}$] is the
Rosseland-averaged opacity (taking into account scattering, per unit mass
of the gas and dust medium).

Equations ~\eqref{therm_sys1}--\eqref{operE} comprise a nonlinear system
of diffusion-type partial differential equations. To solve it, we use a
completely implicit numerical method, in which the right-hand sides of
Eqs.~\eqref{therm_sys1}--\eqref{therm_sys2}, as well as differential
operator \eqref{operE}, depend on the values of the functions on the new
time layer:
\begin{eqnarray}
&&\rho c_{\rm V} \frac{T-T^{n}}{\Delta t} = c \rho \kappa_{\rm P} \left(E_{\rm r} - a T^4\right) + s_{\star}
\label{therm_sys1a}\\
&&\frac{E_{\rm r}-E^{n}_{\rm r}}{\Delta t}  = - c \rho \kappa_{\rm P} \left(E_{\rm r} - a T^4\right) + \hat{\Lambda} E_{\rm r},
\label{therm_sys2a}
\end{eqnarray}
where $T^{n}$ and $E^{n}_{\rm r}$ are the values from the $n$-th time
layer and $T$ and $E_{\rm r}$ are the sought values on the ($n$+1)-th
time layer for a given spatial cell. In the above equations,  for
brevity, we suppressed the superscripts of the ($n$+1)-th time layer of
$\rho,T,\kappa_{\rm P},s_{\star}$ and, $E_{\rm r}$. We also suppressed
the spatial subscripts: for all quantities they correspond to the
considered cell $(i,j)$, except for $s_{\star}$ --- the quantity
connecting three adjacent cells along the radius, and for the operator
connecting the cell $(i,j)$ with four adjacent cells in the radius and
the angle $\theta$. The algorithm for solving this system of equations is
given in Appendix~\ref{AppB}.

\begin{figure*}
\includegraphics[width = 0.5\linewidth]{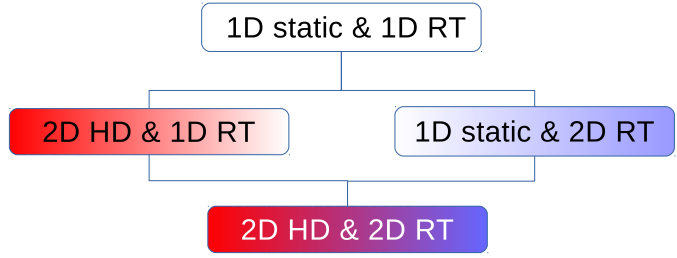}
\caption{Models of the gas and dust disk under consideration: 
(static) hydrostatics, (HD) hydrodynamics, and
(RT) radiative transfer.}
\label{scheme2}
\end{figure*}

The UV radiation intensity required for calculating the heating function
$s_{\star}$ is found for each cell by direct integration of the equation
of radiative transfer from the star to the considered element of the
medium in the radial direction. A fundamental point in calculating  the
function of heating by stellar radiation in our model is considering the
radial gradient of the density inside the cell. In our model, the
function of volume heating by stellar radiation,  $s^{*}$ [\,erg
cm$^{-3}$ s$^{-1}$],  is calculated as proposed
in~\cite{2022ARep...66..321P}:
\begin{equation}
s_{\star}=\rho_{\rm m} \kappa_{\rm UV}
\frac{L\exp{(-\tau)}}{4\pi r_a^2}\, \left(\frac{1-\exp{(-\Delta\tau})}{\Delta\tau} \right),
\label{interpol}
\end{equation}
where $L$ is the star’s luminosity, $\kappa_{\rm UV} = \kappa_{\rm
P}(T_{\star})$ [\,cm$^2$ g$^{-1}$]  is the stellar radiation absorption
coefficient, $r_a$ is the radial distance from the star to the cell’s
inner boundary,  $\tau$ is the total optical thickness along the line of
sight from the star to the cell’s inner boundary, $\Delta \tau = \kappa
\rho_{\rm m} \Delta l$ is the optical thickness of the cell itself along
the beam, $\Delta l$ the beam segment length inside the cell, and
$\rho_{\rm m}=\dfrac{1}{4}\left(\rho_{\rm L}+2\rho_{i}+\rho_{\rm
R}\right)$ is the averaged density along the beam. When deriving
formula~\eqref{interpol} from the formal solution of the radiative
transfer equation, it was assumed that the density along the beam inside
the cell varies linearly from $\rho_{\rm L}$ to $\rho_{i}$ and  from
$\rho_{i}$ to $\rho_{\rm R}$. The values of $\rho_{\rm L}$ and $\rho_{\rm
R}$ at the cell boundaries, in turn, are found using a linear
interpolation of density between the center of the current cell and the
centers of cells adjacent in the radial direction.

The model assumes that the only source of opacity is dust and that the
gas and dust temperatures are equal. The ratio between the dust and gas
densities throughout the disk is assumed to be constant and equal to
0.01; i.e., dust is assumed to be  homogeneously mixed with gas. A
specific feature of the  thermal model is the use of Planck- and
Rosseland-averaged  opacities depending on temperature. These
coefficients  were takes from~\cite{2020ARep...64....1P}, where they are
described in detail. Note that we do not use more realistic  coefficients
that consider, in particular, dust evaporation at high temperatures (see,
e.g., \cite{2003A&A...410..611S}) in order to limit the number of effects
studied.

\subsection{Boundary and Initial Conditions}

As the initial state, we specify a vertically 
hydrostatic Keplerian disk with a temperature of 10\,K and a
distribution of the total ($-\infty < z < \infty$) surface density with a
power-law truncation of the inner boundary as in~\cite{2022ARep...66..321P}:
\begin{equation}
\Sigma(R) = \Sigma_0 \left(1-e^{-\left(\frac{R}{R_0}\right)^{p}} \right)
 \left(\frac{R}{1\,{\rm au}}\right)^{-1},
\label{dens}
\end{equation}
where $\Sigma_0=200$~g cm$^{-2}$ is the normalization of the  surface
density; $R_{0}=3$~au and $p=8$ are the parameters of smoothing the
density distribution near the inner boundary of the disk. The inner and
outer boundaries of the disk are 1 and 20~au. The mass, temperature, and
luminosity of the central star are $M=1M_{\odot}$,  $T_{\star}=6000$\,K,
and $L=5L_{\odot}$ , respectively. At the outer boundary of the
computational domain, we specify a background density of $10^{-19}$~g
cm$^{-3}$ , zero velocity  components, and a background IR radiation
temperature of 10\,K.

\section{Approximate Models of a Gas and Dust Disk}

The aim of this work is to study the role of two- dimensional effects in
the generation of thermal surface  waves. For this, along with the basic
two-dimensional  model, we considered models in which the approximations
that we used in the previously presented  1+1D model
by~\cite{2022ARep...66..321P} are successively replaced with more
rigorous ones. Along with full-valued 2D hydrodynamics  (2D HD), we
considered models in which the gas is in hydrostatic equilibrium in the
$\theta$ direction (1D static). In addition to the 2D approximation in
the transfer of IR radiation (2D RT), we also considered models in which
IR radiation can only propagate in the $\theta$ direction (1D RT).
Figure~\ref{scheme2} shows the combinations  of models we have
considered, and below we briefly describe the implementation of the
approximations used.

Note that the ''1D static \& 1D RT'' combination is most similar in
formulation to the model from our previous
work~\citep{2022ARep...66..321P}. In this article, we consider the ''1D
static \& 1D RT'' model in order to reproduce the surface wave  formation
process obtained in~\cite{2022ARep...66..321P} on a Cartesian grid.

\subsection{Hydrostatic Approximation}

We calculate the disk structure within this approximation 
based on the solution of the following equation:
\begin{equation}
\dfrac{1}{\rho}\frac{d(\rho T)}{d \psi} = -\beta \psi,
\label{eqstatic}
\end{equation}
where $\psi=\pi/2-\theta$ is the angle measured from the equatorial plane
of the disk, $\beta=\dfrac{GM}{R_{\mu}r_{c}}$, $R_{\mu} = \dfrac{k_{\rm
B}}{\mu m_a}$, $k_{\rm B}$ the Boltzmann constant, $\mu=2.3$ is the
average molecular weight, $m_a$ is the atomic mass unit, and $r_c$ is the
radial distance to the cell center. Equation \eqref{eqstatic} can be
derived from the equation for a vertically hydrostatic  disk, assuming
that the ratio $z/r$ is small. However, we use Eq. \eqref{eqstatic} for
all cells of the computational domain. When integrating Eq.
\eqref{eqstatic} numerically, we assume that the temperatures in the
cells are known (determined after calculating the radiative transfer). An
additional condition for integrating Eq.~\eqref{eqstatic} is the
preservation of the mass of the entire $\theta$-column of cells. Thus, in
this approach, we allow the substance to be redistributed in the $\theta$
direction, adjusting to the current thermal structure. Of course, such a
disk will not be hydrostatic in the physical sense, since Eq.
\eqref{eqstatic} is approximate and incorrect as $z/r
\sim 1$. However, it qualitatively correctly reflects the dependence of
the disk height on the thermal structure and is easily solved on a
spherical grid; therefore, it is convenient to use to study the disk
instability.

\begin{figure*}
\includegraphics[width = 0.48\linewidth]{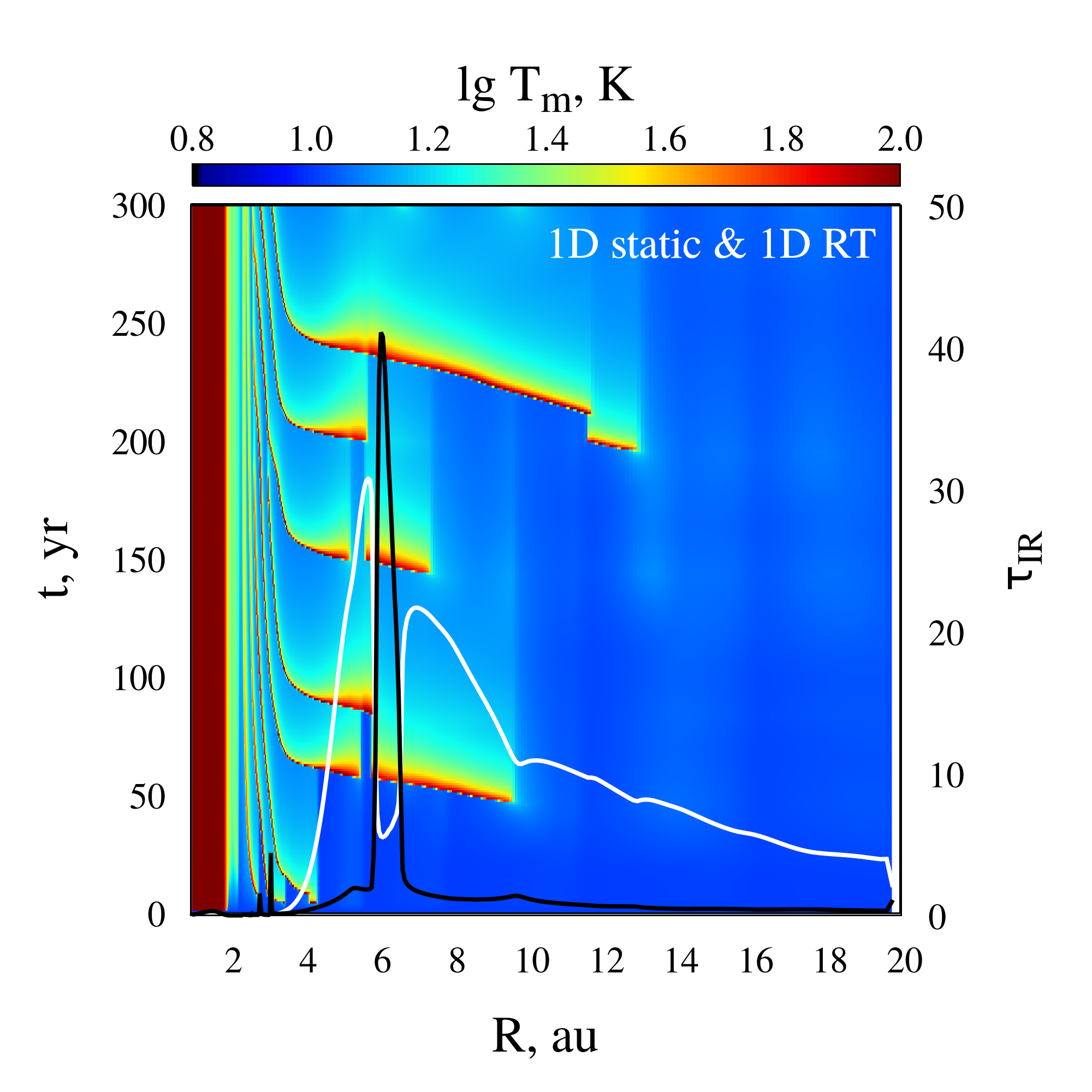}
\includegraphics[width = 0.42\linewidth]{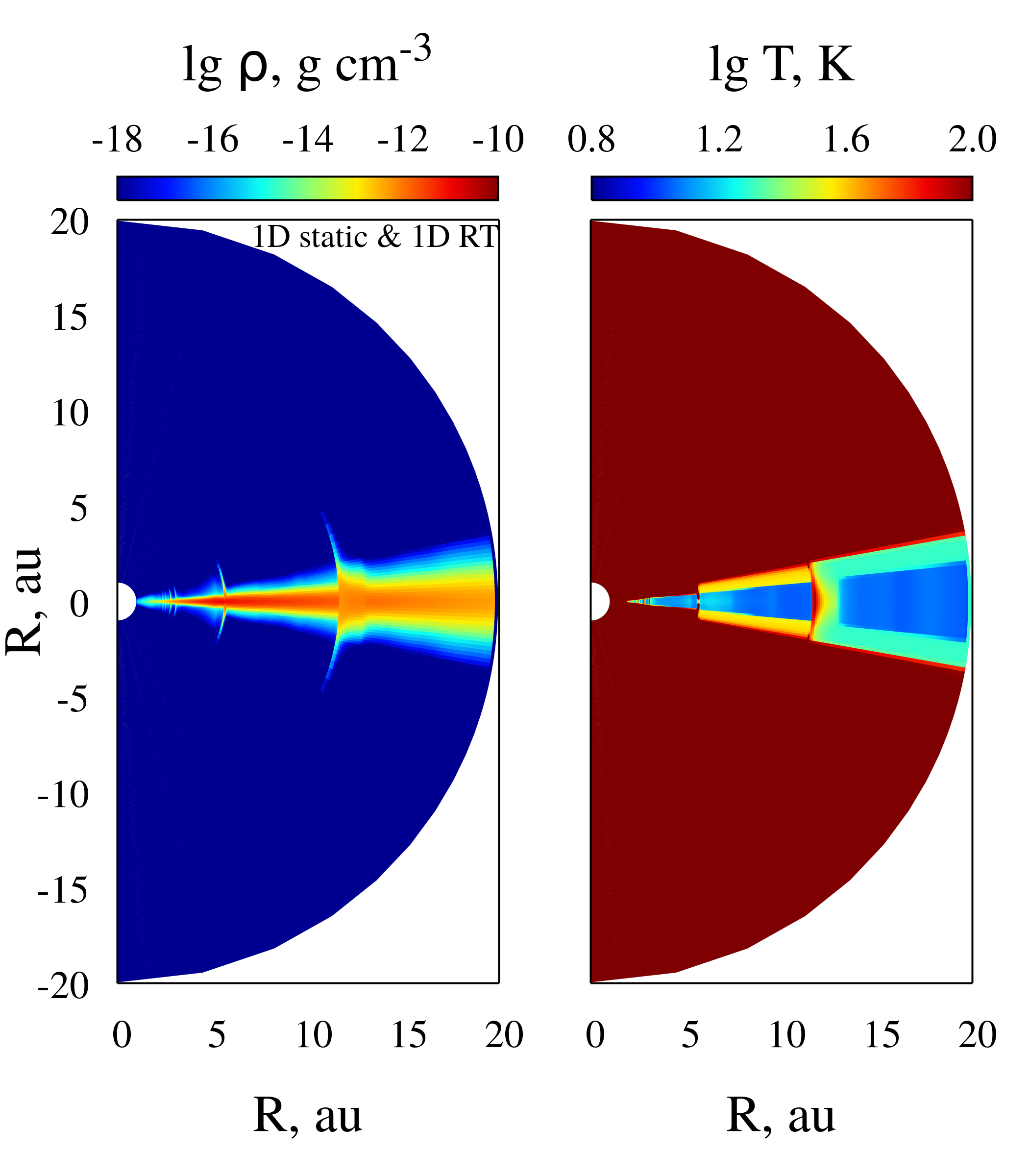}
\caption{Calculation results for the ''1D static \& 1D RT model'': (left)
evolution of equatorial temperature over the first 300~years, (white
line) characteristic thermal time for the entire disk thickness, (black
line) vertical optical thickness to thermal radiation, and (right)
density and temperature distributions in the polar section of the disk at
a time point of 200~years.}
\label{result1}
\end{figure*}

\subsection{One-Dimensional Thermal Model}

In this approximation, we assume that IR radiation propagates only in the
$\theta$ direction, while the UV  radiation of the star heats the disk
only in the radial  direction. Within the equations described in
Section~\ref{sec:base_model} and Appendix~\ref{AppB}, this approximation
is realized by zeroing the fluxes $F_{{\rm r},ra}$ and $F_{{\rm r},rb}$
in expression \eqref{operSF}, after which the determination of the
thermal structure is divided into a number of one-dimensional problems.
The corresponding  finite-difference equations are solved using the
tridiagonal matrix algorithm. In this case, an important point is to
impose appropriate boundary conditions: if we formally take all the cells
with the same $\theta$ in the difference scheme obtained, then the
radiation inside this array will be blocked, since the areas of the cell
faces that are in contact with the polar axis are equal to zero
($S_{\theta a} = 0$ at $\theta=0$  and $S_{\theta b} = 0$ at
$\theta=\pi$). Therefore, we artificially set the radiation energy
density in the polar cells equal to the interstellar  background. Note
that this approximation, like the model of hydrostatic equilibrium with
respect to the angle $\theta$, is, of course, quite rough and cannot be 
considered useful for practical application. We use it only as a tool for
testing the approximations underlying the 1+1D analytical models of
surface wave generation.

\section{Simulation Results}

In this section, we describe the results of numerical simulation of
surface waves, starting with models with the simplest interpretation of
physical processes (hydrostatic and one-dimensional).
Figure~\ref{result1} (left) presents the results of calculating the
evolution of the equatorial disk temperature in the first 300 years using
the ''1D static \& 1D RT model''.

The distribution clearly shows perturbations originating  within 15~au
and propagating from outside to inside. The emergence period and
propagation time of perturbations is $\sim$50~years, which is comparable
to the characteristic thermal time $t_{\rm th} =
\dfrac{3}{8}\dfrac{c_{\rm V}\Sigma^2\kappa_{\rm R}(T_{\rm
m})}{\sigma_{\rm SB} T_{\rm m}^3} $, whose distribution is shown by the
white line taken for the time point of 300~years. In this distribution,
the waves propagating  from outside to inside have discontinuities (see,
e.g., the region in the vicinity of 6~au at a time point of 150~years).
These discontinuities are associated with the formation of an inner hump
on the disk surface, which begins to obscure the currently existing outer
hump. After the inner hump reaches the inner boundary  of the disk, the
outer hump recovers and continues its propagation from the place where it
stopped at the time of the eclipse (the eclipse time is less than the
cooling time of the outer hump).

Figure~\ref{result1} (left) also shows (by the black line) the 
distribution of the optical thickness  $\tau_\text{IR} =
\kappa_\text{P}\Sigma(R)$ in the vertical direction to the disk’s own
thermal radiation at a time point of 300~years. A strong peak in the
vicinity of 6~au is associated with an increase in temperature (and,
accordingly, an increase in $\kappa_\text{P}(T)$) in this region due to
the emerging thermal perturbation. The optical depth exceeds unity in the
region of 4--10~au, which justifies the use of the Eddington
approximation for calculating the diffusion of thermal radiation.

\begin{figure*}
\includegraphics[width = 0.53\linewidth]{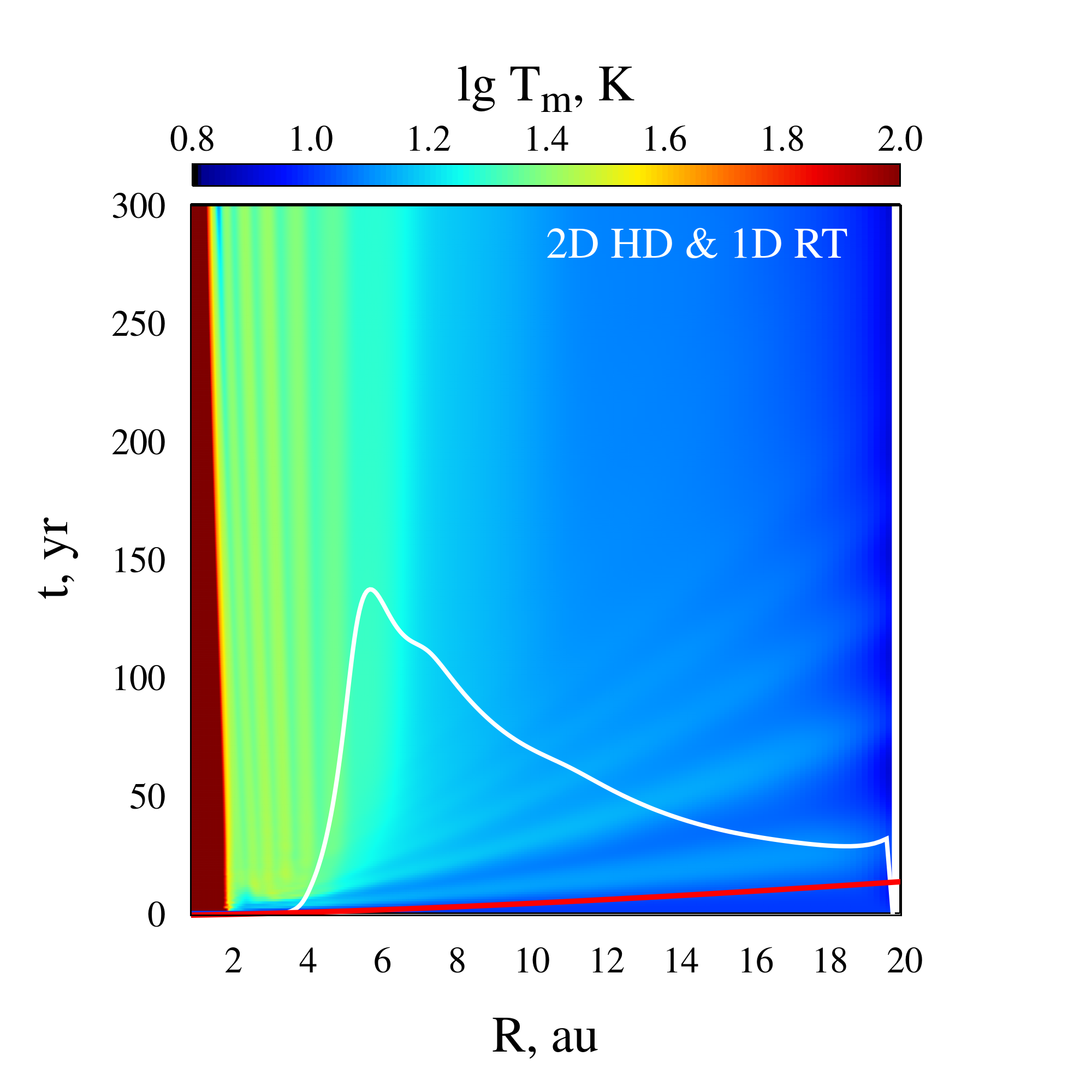}
\includegraphics[width = 0.46\linewidth]{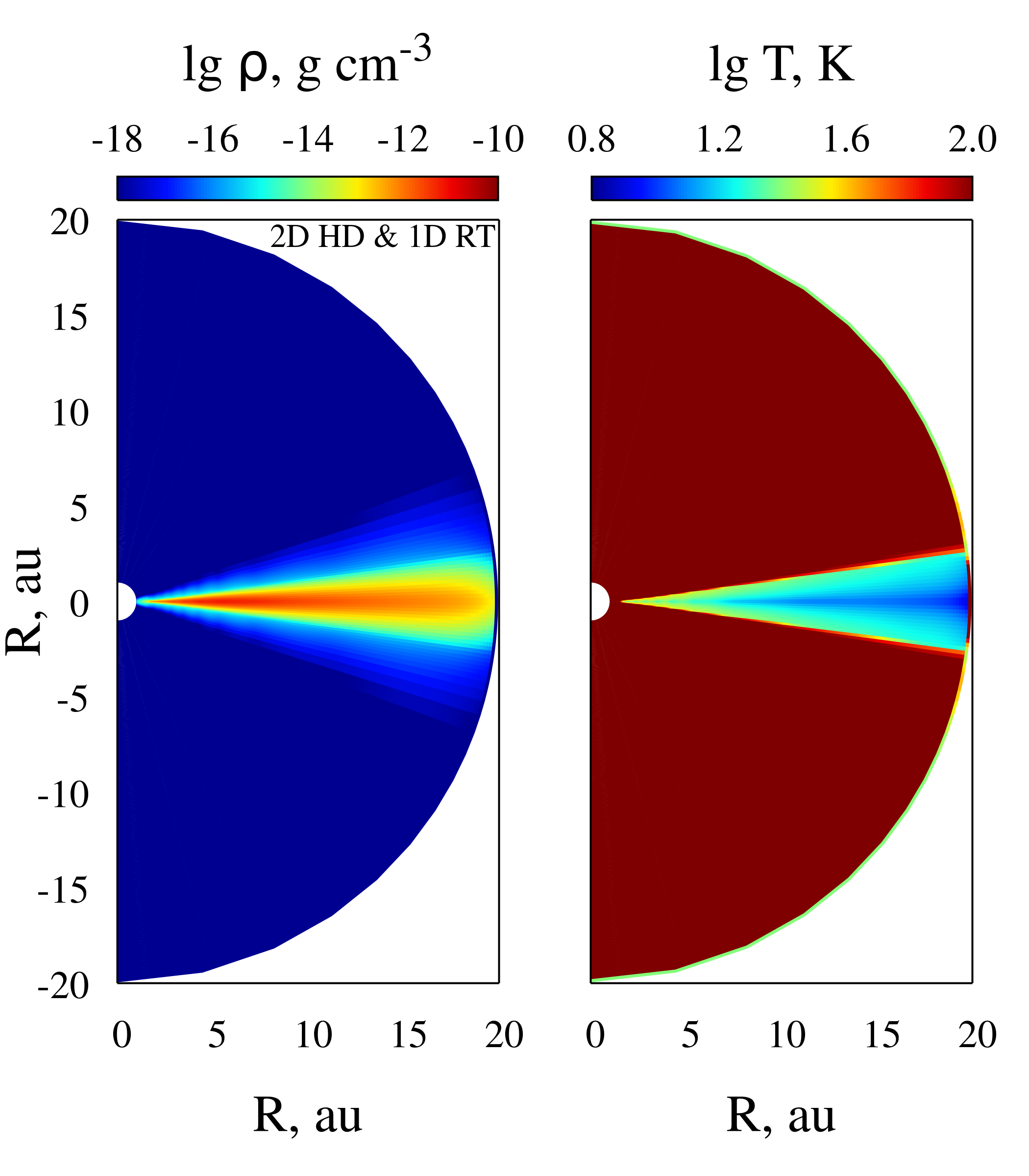}
\caption{Calculation results for the ''2D HD \& 1D RT model'': (left)
equatorial temperature over the first 300~years, (white line)
characteristic thermal time for the entire disk thickness, (red line)
characteristic dynamic (Keplerian) time. The numerical scale for these
times coincides with the scale for evolution time. (Right) Density and
temperature distributions in the polar section of the disk at a time
point of 200~years.}
\label{result2}
\end{figure*}

\begin{figure*}
\includegraphics[width = 0.4\linewidth]{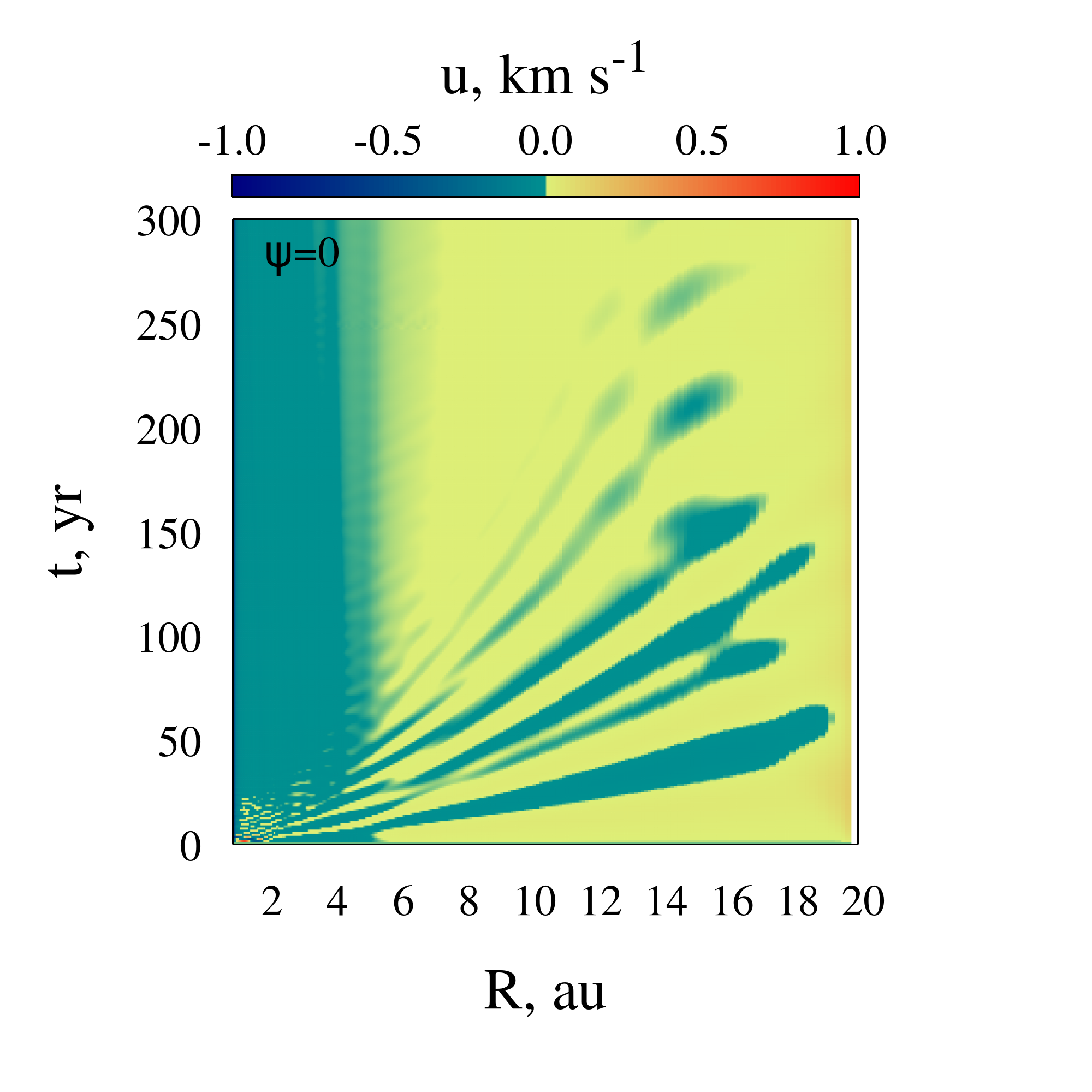}\hspace{-1cm}
\includegraphics[width = 0.4\linewidth]{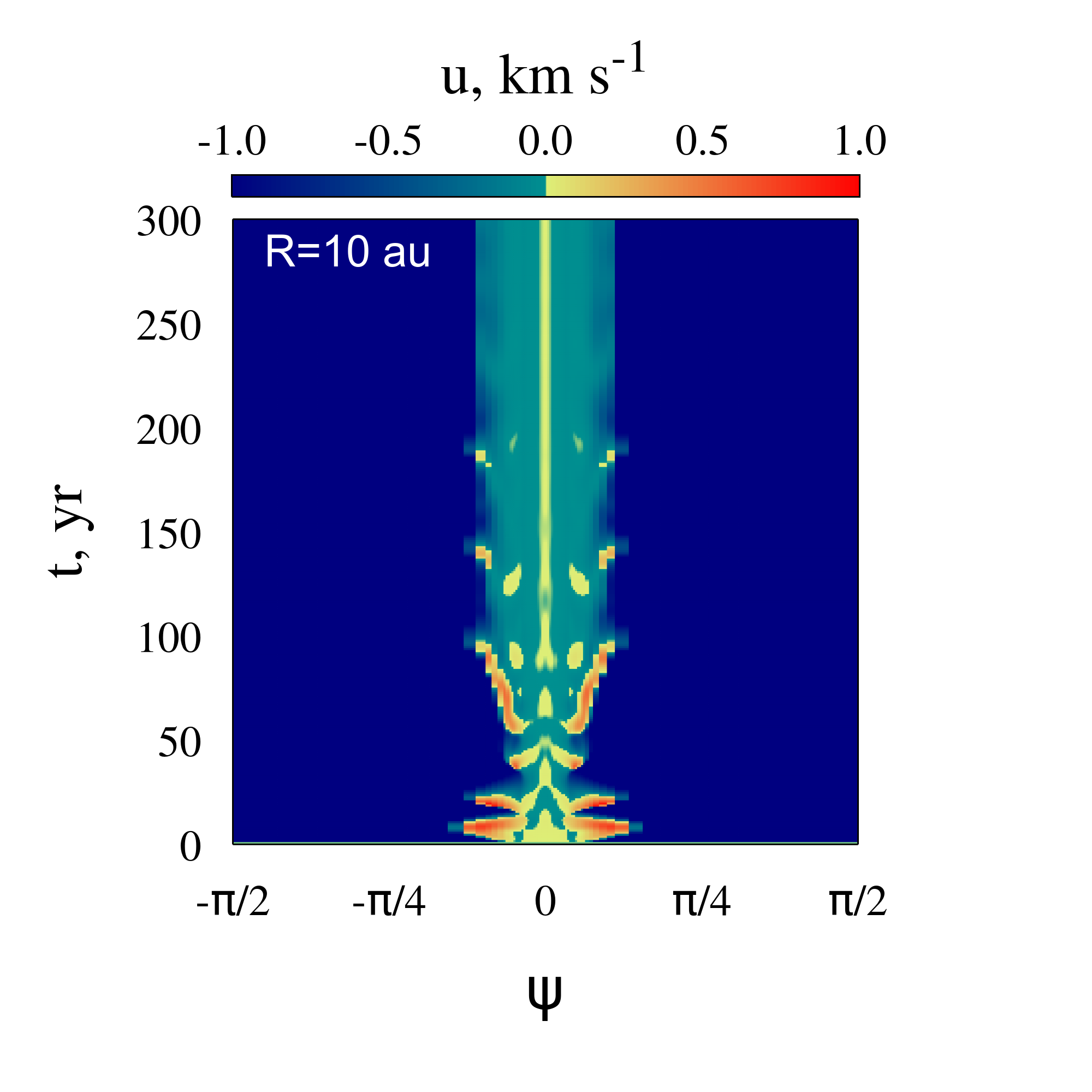}\hspace{-0.5cm}
\includegraphics[width = 0.24\linewidth]{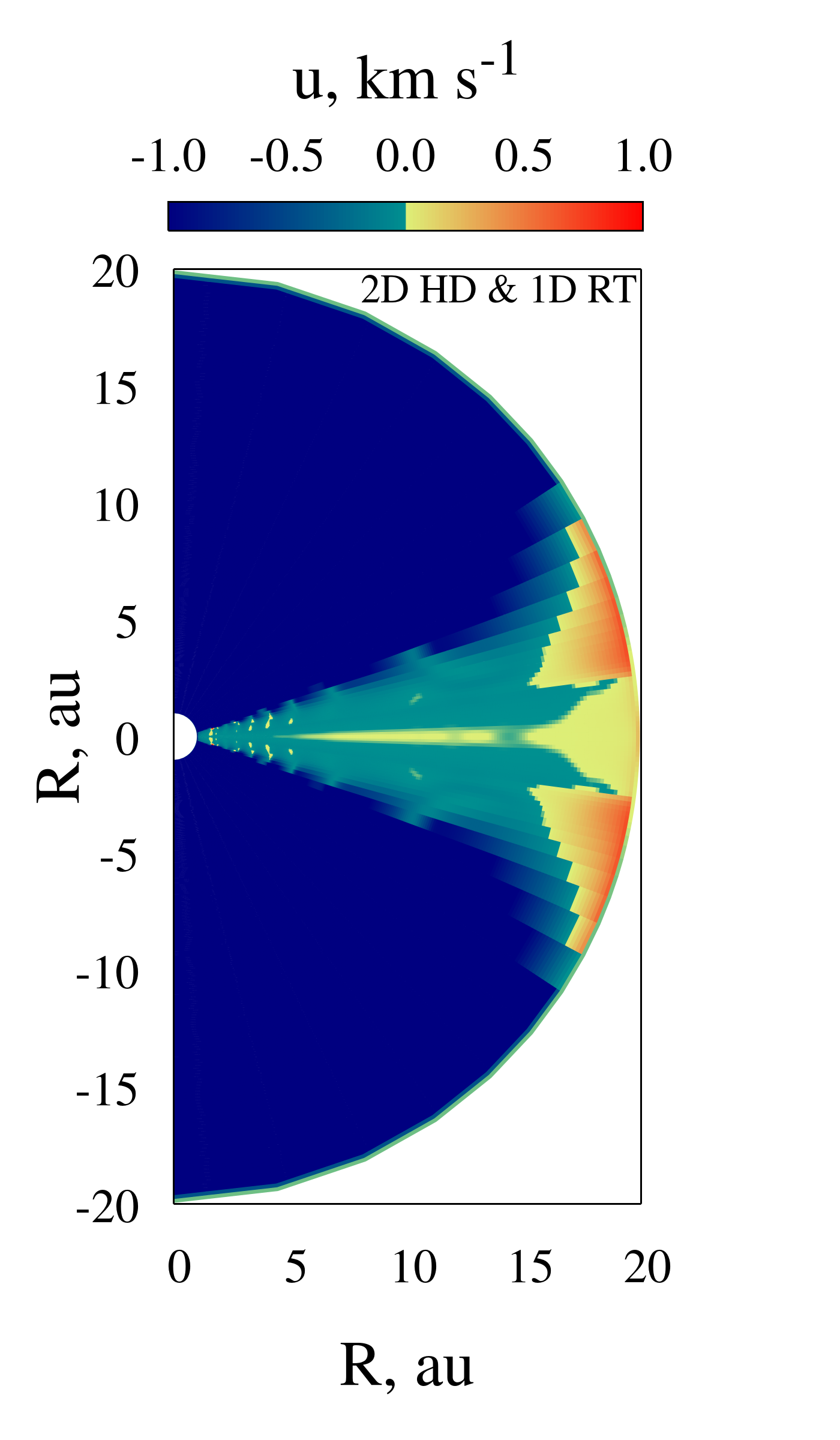}
\caption{Calculation results for the ''2D HD \& 1D RT model'': (left)
evolution of the radial velocity distribution $u(r)$ in the equatorial
plane of the disk; (center) evolution of the latitudinal distribution of
the radial velocity $u(\psi)$, where $\psi$ is the angle measured from
the equator at a radius of 10~au; and (right) distribution of the radial
velocity $u(r,\psi)$ in the polar section of the disk at a time point of
200~years.}
\label{result5}
\end{figure*}

Figure~\ref{result1} (right) shows two-dimensional density and
temperature distributions in the polar section of the disk at a time
point of~200 years. The distributions clearly show two arc-shaped
perturbations in the vicinity of 5 and 12~au. This form of perturbations
is associated with the assumption of the hydrostatic model that the
matter is redistributed only in the $\theta$ direction. For each radial
position in the disk, the temperature decreases from the atmosphere
towards the equator, which is typical of a classical passive disk. The
temperature in the region of the humps, which receive a greater share of
stellar radiation, is increased. The temperature also increases over
$\sim$2~au behind the humps, which is related to the finite time of disk
cooling  due to its own IR radiation. This qualitative picture of disk
evolution is similar to that presented in our  previous
work~\citep{2022ARep...66..321P} for a disk model with similar initial
parameters. Note that this model has a primarily methodological value,
since the unrealistic (arc) shape of traveling perturbations, associated
with a one-dimensional description of the disk structure and radiation
transfer in the $\theta$ direction, does not allow using it in relation
to any observations. At the same time, with the help of this model, we
managed to reproduce the formation of thermal waves within the numerical
code implemented in a spherical coordinate system. This model serves as a
starting point for our subsequent study of two-dimensional effects.

\begin{figure*}
\includegraphics[width = 0.48\linewidth]{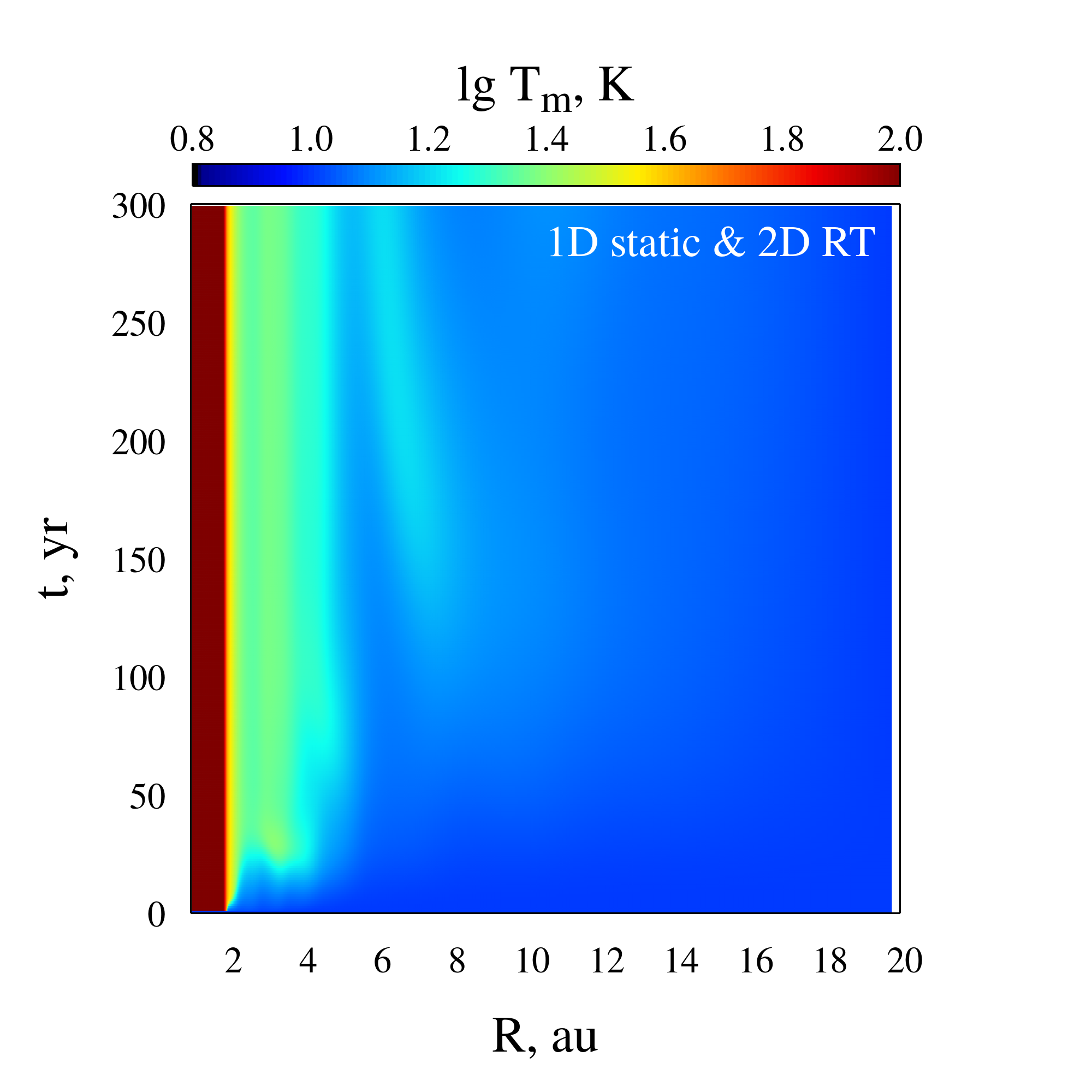}
\includegraphics[width = 0.42\linewidth]{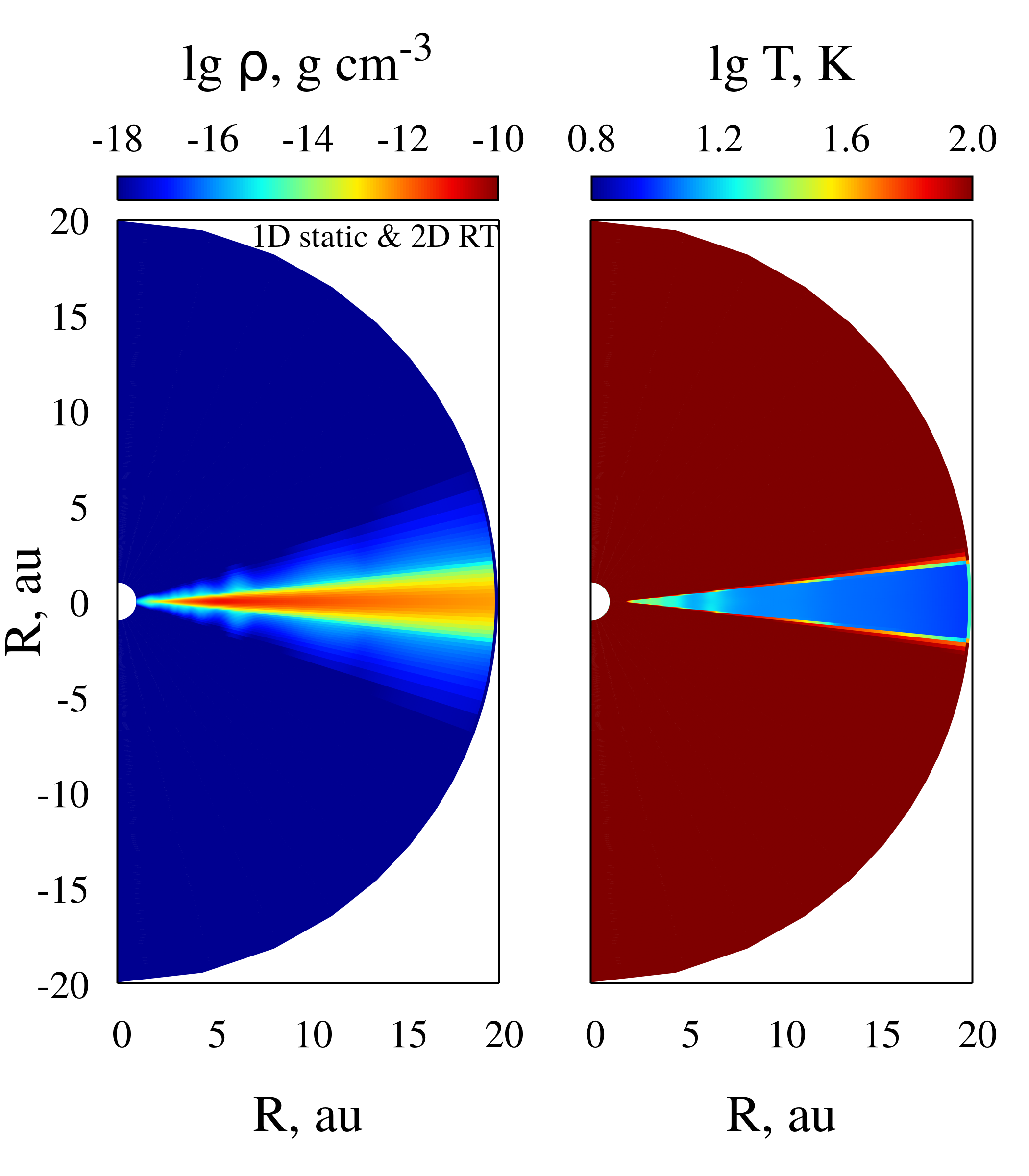}
\caption{Calculation results for the ''1D static \& 2D RT model'': (left)
evolution of equatorial temperature over the first 300~years and (right)
density and temperature distributions in the polar section of the disk at
a time point of 200~years.}
\label{result3}
\end{figure*}

Figure~\ref{result2} shows the results of calculating disk  evolution
within the ''2D HD \& 1D RT model''. Considering the hydrodynamic effects
completely changes the picture of disk evolution. In contrast to the ''1D
static \& 1D RT model'', in which waves periodically  originate in the
region of 6--15~au, perturbations in the ''2D HD \& 1D RT model'' arise
only at the initial times within 4 au and propagate outward. We believe
that these perturbations are associated with a strong hydrodynamic
nonequilibrium of the initial state of the disk. These waves decay with
time and can be traced up to $\sim$150~years. In the inner part of the
disk ($r<5$~au), a stationary annular structure is established.  The
density and temperature distributions in the polar section of the disk
look smooth, except for weak annular perturbations in the inner regions
of the disk. The thermal structure of the disk is standard for a passive
disk: the disk atmosphere is warmer than the equatorial regions and there
is a weak radial temperature  gradient in the equatorial plane.

Figure~\ref{result5} shows the evolution of the radial velocity $u$ in
the equatorial plane, as well as along the angular coordinate for cells
at a radius of 10~au. These distributions  illustrate the propagation and
damping of hydrodynamic  perturbations that arise at the initial moment
of time. Perturbations of radial velocities  $\Delta u\sim 0.2$~km/s are
comparable to sound velocities $c_\text{s}=\left(k_{\rm B}T_\text{m}/(\mu
m_a)\right)^{1/2} \approx 0.19$~km/s  for $T_\text{m}=10$~K. Note that
velocity perturbations in the surface layers ($\psi\approx \pm
0.3$radians) are higher than in the equatorial plane ($\psi = 0$). An
interesting feature of these  distributions is that, over time, a
meridional circulation in the disk is established: in the surface parts
of the disk, matter flows towards the star, while, in the equatorial
region, matter flows outward. Strong velocity  perturbations $\sim$1~km/s
at the outer boundary of the  computational domain are associated with
the boundary conditions imposed.

Thermal waves propagating from outside to inside do not arise in this
model. We believe that the reason is that the emerging surface
perturbations have time to smooth out dynamically before they have time
to  significantly warm the lower layers. Such smoothing is analogous to
the process of relaxation of the initially nonequilibrium state described
above. This assumption  is supported by a comparison of the
characteristic thermal time (duration of heating of the equatorial
layers) and dynamic time $t_{\rm K}=\sqrt{R^3/(GM)}$, shown in the
Fig.~\ref{result2} (left) in white and red respectively. Indeed, in the
region under consideration, the dynamic time is shorter than the thermal
time. It should be noted that the condition of smallness of dynamic time
with respect to thermal time is used as a justification of using vertical
hydrostatic equilibrium, which is the key approximation in the modern
picture of thermal wave formation. However, the results of our simulation
show that this condition leads to dynamic relaxation of perturbations in
the radial direction. Based on this model, we conclude that hydrodynamic
effects can suppress surface waves.

Figure~\ref{result3} shows the results of calculating the disk
evolution using the ''1D static \& 2D RT model''.
The character of disk evolution in this model differs significantly from
the two models considered above. In this model, no periodic traveling
waves arise, but several humps can be distinguished in the disk
structure.  The most noticeable hump is formed in the region of
$\sim$8~au at the time point of $\sim$120~years and slowly moves inward,
but, after 250~years, it becomes almost stationary. The temperature
distribution in the polar section of the disk at a time point of
200~years is more uniform compared to the ''2D HD \& 1D RT model'' and
vertical temperature stratification is not so clearly manifested in it.
We attribute this feature to obscuring the outer layers of the disk by
internal  quasistationary humps; in this case, the thermal structure is
largely determined by the two-dimensional nature of the diffusion of IR
radiation. The general conclusion from this model is that the
two-dimensional transfer of IR radiation rather effectively blurs the
thermal  inhomogeneities of the disk, thereby suppressing (or many times
slowing down compared to the ''1D static \& 1D RT model'') the periodic
formation and propagation of surface waves. At the same time, in this
model,  internal quasi-stationary perturbations are observed that
intercept and process a significant fraction of the  stellar radiation
entering the disk.

\begin{figure*}
\includegraphics[width = 0.48\linewidth]{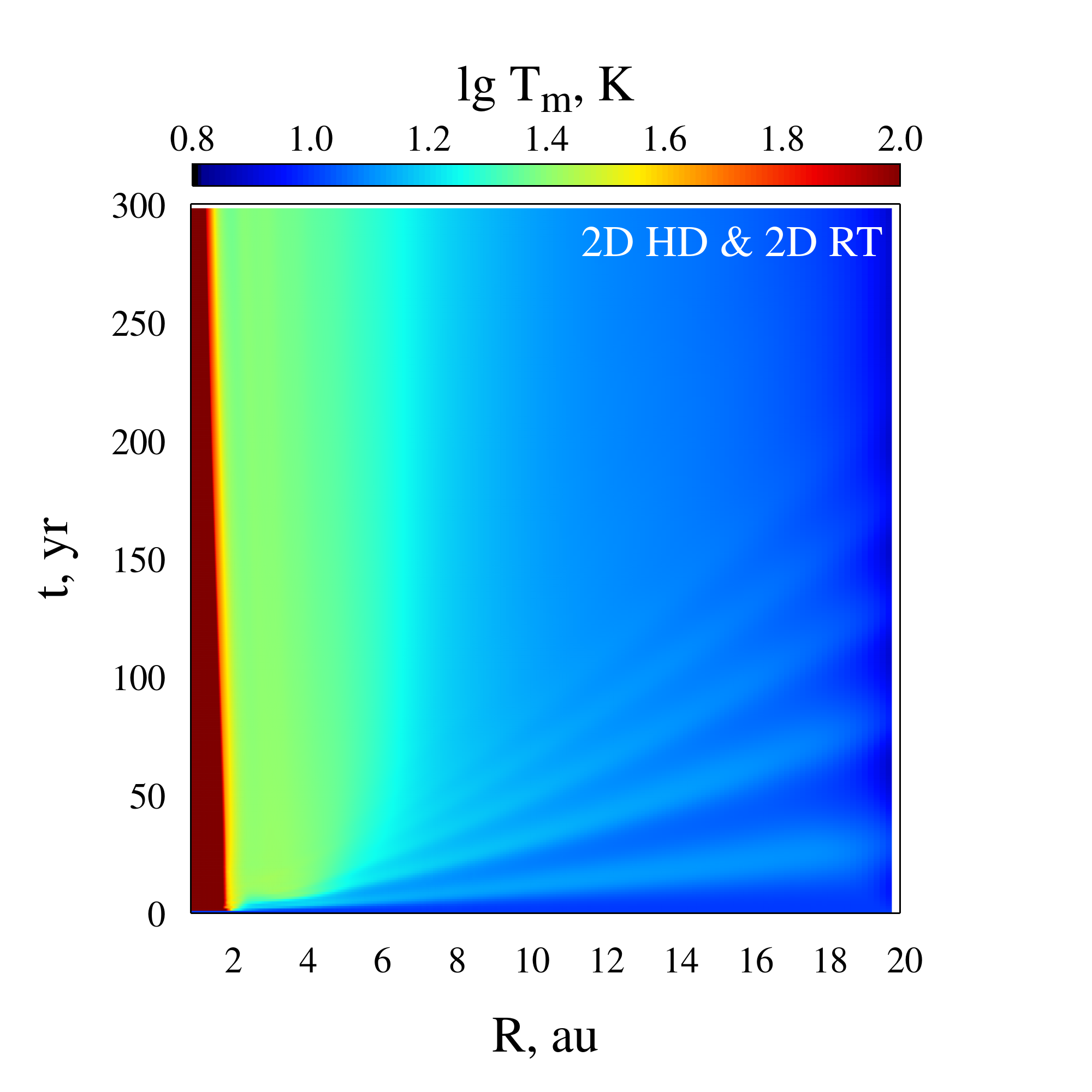}
\includegraphics[width = 0.42\linewidth]{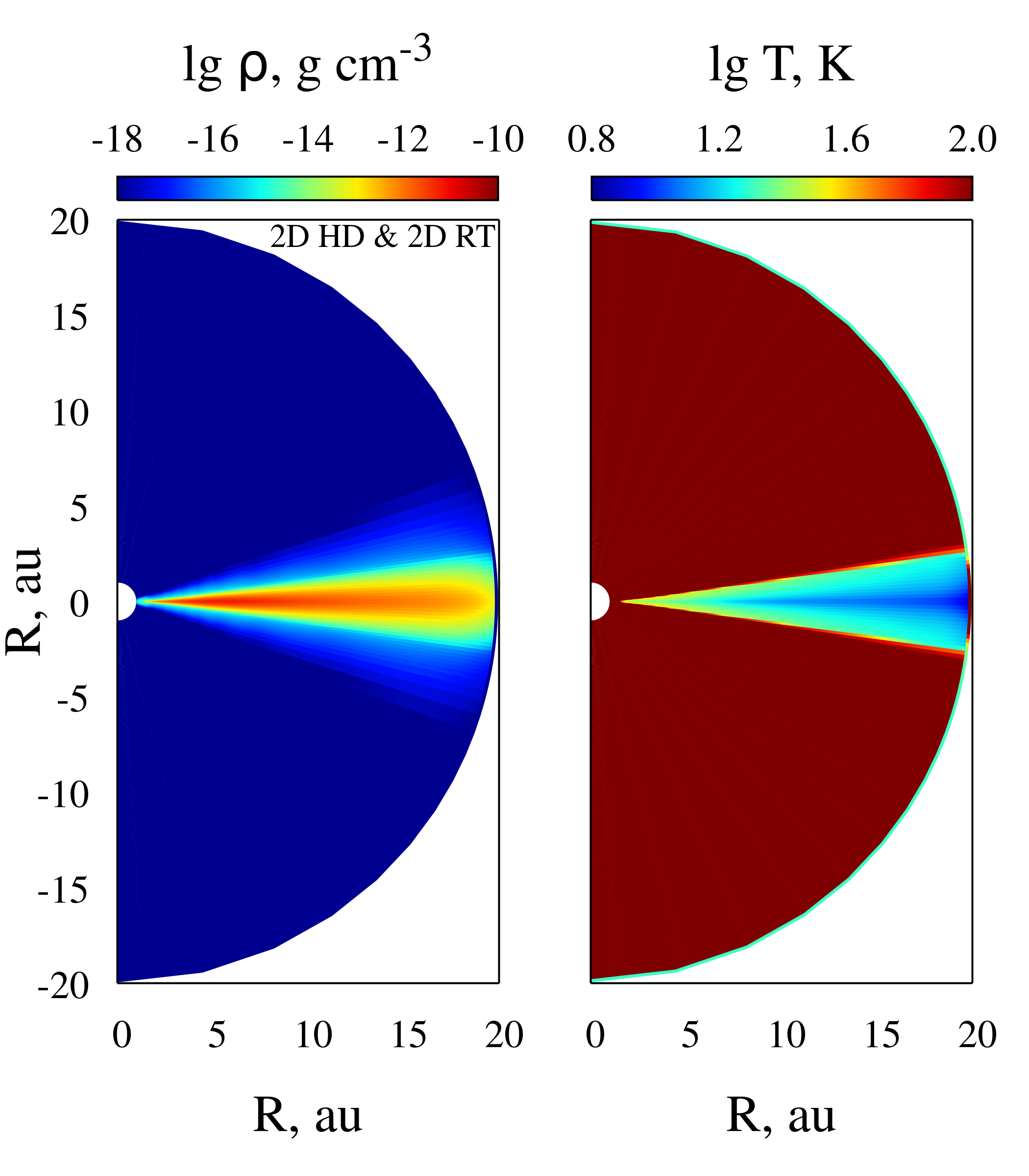}
\caption{Calculation results for the ''2D HD \& 2D RT model'': (left)
evolution of equatorial temperature over the first 300~years and (right)
density and temperature distributions in the polar section of the disk at
a time point of 200~years.}
\label{result4}
\end{figure*}

The conclusions obtained within this model should be compared with the
results of ~\cite{2022arXiv220109241O}, where two- dimensional effects of
radiative transfer are also  considered in the hydrostatic approximation,
but waves there are formed. The key difference between our models is that
the thermal structure of the disk is described in
~\cite{2022arXiv220109241O} in the two-zone approximation, while the
characteristic height of the disk is  determined by the equatorial
temperature. In our model, the hydrostatic structure of the disk is built
in  accordance with the vertical temperature profile.  In addition,
~\cite{2022arXiv220109241O} uses a different method for calculating the
transfer of thermal radiation based on direct integration,  while our
approach uses the diffusion method. As a result, it can be assumed that,
within our model, the relaxation of thermal inhomogeneities is more
efficient, which does not lead to their growth.

Finally, Fig.~\ref{result4} shows the results of calculating the disk
evolution using the ''2D HD \& 2D RT model''.
The dynamics and structure of the disk within this model is close to that
observed in the ''2D HD \& 1D RT model''. The only significant difference
between them is that, in the ''2D HD \& 2D RT model'', the inner region
of the disk ($r<6$~au) has become homogeneous  and is not divided into
rings. The structure of the disk in the ($(r\theta)$) plane corresponds
to a passive disk with pronounced vertical thermal stratification. Based
on the results of this model, a global conclusion can be drawn that the
joint consideration of two-dimensional hydrodynamics and thermal
radiative transfer suppresses  the formation and propagation of surface
thermal  waves in gas and dust disks.

In connection with our results, it is worth noting
the paper by~\cite{2022ApJ...936...16M}, the preprint of which appeared at the
stage of reviewing our work. \cite{2022ApJ...936...16M}
independently of us, using their numerical model, also came to the
conclusion that thermal waves are suppressed if hydrodynamic effects
and nonstationary character of heat transfer are considered.

\section{Conclusions}

This work is a logical continuation of the research presented in our
article~\cite{2022ARep...66..321P}. The main objective of this work was
to study surface thermal waves in gas and dust disks with a more
realistic description of the  processes, considering two-dimensional
hydrodynamic effects and two-dimensional effects of thermal  radiative
transfer (in the $(r\theta)$ plane). To do this, we have developed a
model of the evolution of an axisymmetric disk, destined to successively
eliminate the  approximations that underlie the modern theory of thermal
surface waves. In this two-dimensional numerical code, we also
implemented simplified approaches used in 1+1D models: the approximation
of vertical hydrostatic equilibrium and vertical diffusion of IR
radiation, and reproduced the formation of traveling surface waves
from~\cite{2022ARep...66..321P}. Then we showed that  replacing the
hydrostatic equilibrium approximation by hydrodynamic simulation leads to
the disappearance of thermal surface waves. The same result is obtained
when the one-dimensional approximation for  calculating thermal radiation
is replaces with two-dimensional  simulation. Our results indicate that
the jointly taking into account two-dimensional effects associated  with
hydrodynamics and thermal radiative transfer  suppresses the development
of thermal surface waves, which arise within the 1+1D approach.

The conclusion drawn here must be taken with some reservation due to the
limitations of our model. The method we used for calculating the
radiative transfer is based on the Eddington approximation, which assumes
that the IR radiation field is isotropic. This approximation exaggerates
the diffusion nature of radiation propagation in optically thin and
transition regions and, therefore, leads to a more efficient  smearing of
inhomogeneities. It is necessary to verify the conclusions using more
correct methods for calculating  the radiative transfer, such as flux
limited diffusion method \citep{1981ApJ...248..321L} or the method of the
variable Eddington tensor \citep{1992ApJS...80..819S}. It was noted 
in~\cite{2022arXiv220109241O} that the development of instability can
also be suppressed by an  insufficiently high spatial resolution of the
numerical model. In our calculations, due to high computational costs, we
were limited to a relatively sparse spatial grid ($N_r \times N_{\theta}
= 360\times64$, but it is nonuniform). Therefore,  further analysis on
finer grids is necessary. It should be noted that our conclusions cannot
state that the development of instability is fundamentally impossible.
Our results only show the absence of 2D surface waves under specific
physical conditions, under which they are spontaneously formed in the
1+1D model. This problem should be investigated for a wide range of
parameters of gas and dust disks, in particular, when the characteristic
thermal and dynamic times are comparable.

\appendix
\section{Flux Vector Components for the Finite-Difference Equation of Hydrodynamics}
\label{AppA}

The components of the flux vectors entering into
Eq.~\eqref{advection} for a finite-difference grid 
in spherical coordinates, taking into account the axial symmetry of the
problem, have the following form:
\begin{align*}
&\Delta F_{1}=F_{1a} - F_{1b} \\
&\Delta F_{2}=F_{2a} - F_{2b} \\
&\Delta F_{3}=F_{3a} - F_{3b} \\
&\Delta F_{4}=F_{4a} - F_{4b} \\
&\Delta F_{5}=F_{5a} - F_{5b} \\
\\
&\Delta G_{1}=G_{1a} - G_{1b} \\
&\Delta G_{2}=G_{2a}\cos{\theta} - G_{2b}\cos{\theta}
             +G_{3a}\sin{\theta} + G_{3b}\sin{\theta} \\
&\Delta G_{3}=-G_{2a}\sin{\theta} - G_{2b}\sin{\theta}
              +G_{3a}\cos{\theta} - G_{3b}\cos{\theta} \\
&\Delta G_{4}=G_{4a} - G_{4b} \\
&\Delta G_{5}=G_{5a} - G_{5b} \\
\\
&\Delta H_{1}=0 \\
&\Delta H_{2}=H_{4a}\sin{\varphi}\sin{\theta}
             +H_{4b}\sin{\varphi}\sin{\theta} \\
&\Delta H_{3}=H_{4a}\sin{\varphi}\cos{\theta}
             +H_{4b}\sin{\varphi}\cos{\theta} \\
&\Delta H_{4}=-H_{2a}\sin{\varphi}\sin{\theta}
              -H_{2b}\sin{\varphi}\sin{\theta}\\
&\hspace{2cm} -H_{3a}\sin{\varphi}\cos{\theta}
              -H_{3b}\sin{\varphi}\cos{\theta}\\
&\Delta H_{5}=0,
\end{align*}
where
\begin{align*}
&  F_{1a} = (\tilde{\rho} \tilde{u})_{ra}\, S_{ra}
&& F_{1b} = (\tilde{\rho} \tilde{u})_{rb}\, S_{rb} \\
&  F_{2a} = (\tilde{\rho} \tilde{u}^2 + \tilde{P})_{ra}\, S_{ra}
&& F_{2b} = (\tilde{\rho} \tilde{u}^2 + \tilde{P})_{rb}\, S_{rb} \\
&  F_{3a} = (\tilde{\rho} \tilde{u} \tilde{v})_{ra}\, S_{ra}
&& F_{3b} = (\tilde{\rho} \tilde{u} \tilde{v})_{rb}\, S_{rb} \\
&  F_{4a} = (\tilde{\rho} \tilde{u} \tilde{w})_{ra}\, S_{ra}
&& F_{4b} = (\tilde{\rho} \tilde{u} \tilde{w})_{rb}\, S_{rb} \\
&  F_{5a} = (\tilde{E} + \tilde{P} \tilde{u})_{ra}\, S_{ra}
&& F_{5b} = (\tilde{E} + \tilde{P} \tilde{u})_{rb}\, S_{rb} \\
\\
 & G_{1a} = (\tilde{\rho} \tilde{v})_{\theta a}\, S_{\theta a}
&& G_{1b} = (\tilde{\rho} \tilde{v})_{\theta b}\, S_{\theta b} \\
 & G_{2a} = (\tilde{\rho} \tilde{v} \tilde{u})_{ra}\, S_{\theta a}
&& G_{2b} = (\tilde{\rho} \tilde{v} \tilde{u})_{rb}\, S_{\theta b} \\
 & G_{3a} = (\tilde{\rho} \tilde{v}^2 + \tilde{P})_{\theta a}\, S_{\theta a}
&& G_{3b} = (\tilde{\rho} \tilde{v}^2 + \tilde{P})_{\theta b}\, S_{\theta b} \\
 & G_{4a} = (\tilde{\rho} \tilde{v}  \tilde{w})_{\theta a}\, S_{\theta a}
&& G_{4b} = (\tilde{\rho} \tilde{v} \tilde{w})_{\theta b}\, S_{\theta b} \\
 & G_{5a} = (\tilde{E} + \tilde{P}\tilde{v})_{\theta a}\, S_{\theta a}
&& G_{5b} = (\tilde{E} + \tilde{P}\tilde{v})_{\theta b}\, S_{\theta b} \\
\\
 & H_{2a} = (\tilde{\rho} \tilde{w} \tilde{u})_{\varphi a}\, S_{\varphi a}
&& H_{2b} = (\tilde{\rho} \tilde{w} \tilde{u})_{\varphi b}\, S_{\varphi b} \\
 & H_{3a} = (\tilde{\rho} \tilde{w} \tilde{v})_{\varphi a}\, S_{\varphi a}
&& H_{3b} = (\tilde{\rho} \tilde{w} \tilde{v})_{\varphi b}\, S_{\varphi b}\\
 & H_{4a} = (\tilde{\rho} \tilde{w}^2 + \tilde{P})_{\varphi a}\, S_{\varphi a}
&& H_{4b} = (\tilde{\rho} \tilde{w}^2 + \tilde{P})_{\varphi b}\, S_{\varphi b}
\end{align*}
In the above expressions, the tilde denotes the  quantities found from
the solution of the problem of the decay of an arbitrary discontinuity
for the corresponding  cell’s face, marked with a subscript; $S_{ra}$,
$S_{rb}$, $S_{\theta a}$, $S_{\theta b}$, $S_{\varphi a}$, and
$S_{\varphi b}$ are the areas of the cell’s faces (see
Fig.~\ref{scheme1}). Note that the trigonometric functions and the mixing
of the flux components in the above expressions  for the components of
$\Delta G$ and $\Delta H$ are associated  with the transformation of the
local basis (and, accordingly, the coordinates of the velocity vectors)
upon the transition between cells in $\theta$ and $\varphi$. At the same
time, zero values for $\Delta H_1$ and $\Delta H_5$ are obtained taking
into account the assumed axial symmetry of the problem.

\section{Method for Solving the System of Thermal Radiative Transfer Equations}
\label{AppB}

The system of equations for the thermal evolution
of the medium \eqref{therm_sys1a}--\eqref{therm_sys2a} is solved using Newton’s
iterations; for this, the equations are linearized using
the approximation
\begin{equation*}
T^4 \approx 4 T_k^3 T - 3T_k^4,
\end{equation*}
where $T_k$ is the temperature value at the previous ($k$-th)
iteration, after which Eqs.~\eqref{therm_sys1a}--\eqref{therm_sys2a} can be
reduced to
\begin{eqnarray}
&&T=\frac{b_{\rm d}+\omega_{\rm p}\Delta t\, E_{\rm r}}{c_{\rm d}+c_{\rm r}}
\label{therm_sys1b}\\
&& \left[1+\frac{c_{\rm d}}{c_{\rm d}+c_{\rm r}}\omega_{\rm p} \Delta t - \Delta t\, \hat{\Lambda}\right] E_{\rm r} = g.
\label{therm_sys2b}
\end{eqnarray}
The coefficients $c_{\rm d}$, $\omega_{\rm p}$, $c_{\rm r}$, $b_{\rm d}$, and $g$ in these equations
are calculated as follows:
\begin{eqnarray*}
&& c_{\rm d} = \rho c_{\rm V} \\
&& \omega_{\rm p} = c \rho \kappa_{\rm P} \\
&& c_{\rm r} = 4aT_{k}^3 \omega_{\rm p} \Delta t \\
&& b_{\rm d} = c_{\rm d}T^{n} + \frac{3}{4} c_{\rm r} T_{k} +\rho S \Delta t\\
&& g = E^{n}_{\rm r}-\frac{3}{4}c_{\rm r}T_{k}+\frac{c_{\rm r}b_{\rm d}}{c_{\rm d}+c_{\rm r}}.
\end{eqnarray*}
We approximate the differential operator $\hat{\Lambda}$ in 
spherical coordinates in the following finite difference form:
\begin{equation}
\hat{\Lambda} E_{\rm r} = \frac{1}{ \Delta V}\left(
S_{ra}F_{{\rm r,}ra} - S_{rb}F_{{\rm r,}rb} +
S_{\theta a}F_{{\rm r,}\theta a} - S_{\theta b}F_{{\rm r,}\theta b}
\right),
\label{operSF}
\end{equation}
where $\Delta V$ is the cell volume and $S_{ra}$, $S_{rb}$, $S_{\theta a}$, and
$S_{\theta b}$
are the current cell faces’ areas (see Fig.~\ref{scheme1}). The fluxes
through the cell’s faces are found by the formulas
\begin{eqnarray*}
&& F_{{\rm r,}ra}=-\frac{1}{\sigma_{ra}}\frac{E_{{\rm r}}(i,j)-E_{{\rm r}}(i-1,j)}{\Delta r_{a}}\\
&& F_{{\rm r,}rb}=-\frac{1}{\sigma_{rb}}\frac{E_{{\rm r}}(i+1,j)-E_{{\rm r}}(i,j)}{\Delta r_{b}}\\
&& F_{{\rm r,}\theta a}=-\frac{1}{\sigma_{\theta a}}\frac{E_{{\rm r}}(i,j)-E_{{\rm r}}(i,j-1)}{r_{c} \Delta \theta_{a}}\\
&& F_{{\rm r,}\theta b}=-\frac{1}{\sigma_{\theta b}}\frac{E_{{\rm r}}(i,j+1)-E_{{\rm r}}(i,j)}{r_{c} \Delta \theta_{b}},
\end{eqnarray*}
where $E_{\rm r}(i,j)$ is the energy in the current cell with
indices $(i,j)$, $r_{c}$~is the radial coordinate of the center of
the current cell, the values of $\sigma=3\rho\kappa_{\rm
R}/c$ are calculated 
based on the gas density and medium temperature 
for the corresponding faces by interpolating the
central values, and $\Delta r$ and $r \Delta \theta$ are the distances from
the center of the current cell to the centers of adjacent
cells for the corresponding faces (see Fig.~\ref{scheme1}). With this,
Eq. \eqref{therm_sys2b} can be rewritten in the following operator
form:
\begin{equation}
\hat{\Omega} E_{\rm r} = {\bf g},
\label{ALU}
\end{equation}
where $\hat{\Omega} = \left(1+\dfrac{c_{\rm d}}{c_{\rm d}+c_{\rm
r}}\omega_{\rm p} \Delta t\right)\hat{I} - \Delta t\, \hat{\Lambda}$,
and $\hat{I}$ is the unit tensor. Equation \eqref{ALU} is a compact representation
of a system of linear algebraic equations with a sparse
five-diagonal matrix. We solve this system of equations 
using the GMRES method~\citep{doi:10.1137/0907058} or the alternating 
direction implicit (ADI) method~\citep{Samarskii:1978}, depending
on the complexity of the problem. Our numerical
experiments with this model show that the GMRES
method has good stability, but is several times slower
than the ADI method. For the ADI method, the
choice of iterative parameters is important, which we
find based on the closeness of the results to the solution 
of the system by the GMRES method.

\section*{Acknowledgments}
We are grateful to the referee for valuable comments and
suggestions for improving the article.

\section*{Funding}
This work was supported by the Russian Foundation for
Basic Research (project no. 20-32-90103). 
V. V. Akimkin acknowledges the support by the Theoretical 
Physics and Mathematics Advancement Foundation “BASIS” (grant no.20-1-2-20-1).

\section*{Conflict of Interest}
The authors declare that they have no conflicts of interest.

\bibliographystyle{mn2e}
\bibliography{twave}

\end{document}